\documentclass[]{svmult}

\usepackage{mathptmx}       
\usepackage{helvet}         
\usepackage{courier}        
\usepackage{type1cm}        
\usepackage{makeidx}         
\usepackage{graphicx}        
\usepackage{multicol}        
\usepackage[bottom]{footmisc}
\usepackage{cleveref}
\usepackage{glossaries}
\usepackage{color}
\usepackage{amssymb}
\usepackage{bm}
\usepackage{soul}
\usepackage{natbib} 

\newcommand{\newacronymtwo}[3]{%
 \newcommand{#1}[1]{#3##1 (#2##1)%
 \renewcommand{#1}[1]{#2####1}}%
}

\newacronymtwo{\NS}{NS}{neutron star}
\newacronymtwo{\PNS}{PNS}{proto-neutron star}
\newacronymtwo{\CCSN}{CCSN}{core collapse supernova}
\newacronymtwo{\SN}{SN}{supernova}
\newacronymtwo{\BH}{BH}{black hole}
\newacronymtwo{\EOS}{EoS}{equation of state}

\newcommand{\apjl}{\textrm{Astrophys. J. Lett.}}
\newcommand{\apj}{\textrm{Astrophys. J. }}
\newcommand{\apjs}{\textrm{Astrophys. J. Sup.}}
\newcommand{\mnras}{\textrm{Mon. Not. Roy. Astro. Soc.}}
\newcommand{\aap}{\textrm{Astron. \& Astrophys.}}
\newcommand{\aaps}{\textrm{Astron. \& Astrophys., Sup.}}
\newcommand{\physrep}{\textrm{Phys. Rep.}}
\newcommand{\prd}{\textrm{Phys. Rev. D}}
\newcommand{\prc}{\textrm{Phys. Rev. C}}
\newcommand{\iaucirc}{\textrm{IAU Circular}}

\newcommand{\td}[2]{\frac{d #1}{d #2}}
\newcommand{\thd}[3]{\left(\frac{\partial #1}{\partial #2}\right)_{#3}}

\begin{document}
\title*{Neutrino Signatures From Young Neutron Stars}
\author{Luke F. Roberts and Sanjay Reddy}
\institute{Luke F. Roberts \at NASA Einstein Fellow, TAPIR, California 
Institute of Technology, Pasadena, CA. \email{lroberts@tapir.caltech.edu}
\and Sanjay Reddy \at Institute for Nuclear Theory, University of Washington,
Seattle, WA. \email{sareddy@uw.edu} \and INT-PUB-16-051}
%
%
\maketitle


\abstract{After a successful \CCSN{} explosion, a hot dense \PNS{} is left as a
remnant.  Over a time of twenty or so seconds, this \PNS{} emits the majority of
the neutrinos that come from the \CCSN{}, contracts, and loses most of its lepton
number.  This is the process by which all neutron stars in our galaxy are likely
born.  The emitted neutrinos were detected from SN 1987A and they will be
detected in much greater numbers from any future galactic \CCSN{}.  These
detections can provide a direct window into the properties of the dense matter
encountered inside neutron stars, and they can affect
nucleosynthesis in the material ejected during the \CCSN{}.  In this chapter, we
review the basic physics of \PNS{} cooling, including the basic equations of
\PNS{} structure and neutrino diffusion in dense matter.  We then discuss how
the nuclear equation of state, neutrino opacities in dense matter, and
convection can shape the temporal behavior of the neutrino signal.  We also
discuss what was learned from the late time SN 1987A neutrinos, the prospects for
detection of these neutrinos from future galactic \CCSN{e}, and the effects
these neutrinos can have on nucleosynthesis.}


\section{Introduction}
A nascent \NS{} is often left as the remnant of a successful \CCSN{}.  This young
\NS{} emits a copious number of neutrinos over the first few seconds of its
life.  During this time it is referred to as a \PNS{}.  Due to the
high densities and temperatures encountered inside the \PNS{}, neutrinos
cannot freely escape but instead must diffuse out over a period of about a
minute \citep{Burrows:86}.  This neutrino emission is powered by a large fraction of the
gravitational binding energy released by taking the iron core of a massive star
and transforming it into a \NS{} ($2-5 \times10^{53}$ ergs) \citep{Baade:34}.
After about a minute, neutrinos can escape freely, which demarcates the
transition from \PNS{} to \NS{}.  This qualitative picture of late \PNS{}
neutrino emission was confirmed when about thirty neutrinos were observed from
\SN{} 1987A over a period of about fifteen seconds \citep{Bionta:87, Hirata:87}.  If a \CCSN{} were observed in
our galaxy today, modern neutrino detectors would see thousands of events
\citep{Scholberg:12}. The neutrino signal is shaped by the nuclear \EOS{} and
neutrino opacities.  Therefore, detection of galactic \CCSN{} neutrinos would
give a detailed window into the birth of \NS{s} and the properties of matter at
and above nuclear density.   


In addition to direct neutrino detection, there are other reasons why
understanding the properties of these late-time \CCSN{} neutrinos is important.
First, they can influence nucleosynthesis in \CCSN{e} \citep{Woosley:90}.  In
particular, \PNS{} neutrino emission almost wholly determines what nuclei are
synthesized in baryonic material blown off the surface of \PNS{s}
\citep{Woosley:94, Hoffman:97, Roberts:10}.  Second, the integrated neutrino
emission from \CCSN{e} receives a large contribution from \PNS{} neutrinos.
Therefore, accurate models of \PNS{} neutrino emission can contribute to
understanding the diffuse \SN{} neutrino background \citep{Nakazato:15}.
Finally, the neutrino emission from the ``neutrinosphere'' of \PNS{s} gives the
initial conditions for the study of both matter-induced and neutrino-induced
neutrino oscillations \citep{Duan:06}.  The rate of \PNS{} cooling also
has the potential to put limits on exotic physics and possible extensions of the
standard model using data already in hand from \SN{} 1987A \citep{Keil:97,
Pons:01a, Pons:01}.



In this chapter, we discuss \PNS{} cooling and the late time \CCSN{}
neutrino signal.  In section \ref{sec:pns_cooling},
we focus on the basic equations of \PNS{} cooling (section \ref{sec:pns_cooling_eq})
and models of \PNS{} cooling (sections \ref{sec:pns_evolution} and
\ref{sec:pns_neutrino_emission}).  In sections \ref{sec:eos}, we discuss the various
ingredients that shape the \CCSN{} neutrino signal, the nuclear equation of
state, neutrino opacities, and convection, respectively.  Finally--in sections
\ref{sec:1987a_neutrinos}, \ref{sec:galactic_sn_neutrinos}, and
\ref{sec:nucsyn}--we discuss the observable consequences of late time \CCSN{}
neutrinos.  Throughout the article, we set $\hbar = c = 1$. 

\section{\PNS{} Cooling}
\label{sec:pns_cooling}

Essentially, all of the energy that powers the neutrino emission during a
\CCSN{} comes from the gravitational binding energy released when taking the
white dwarf like iron core of the massive progenitor star and turning it into a
\NS{} \citep{Baade:34}, which is
\begin{equation}
E_{\rm SN} \sim \frac{3G M_{\rm pns}^2}{5 r_{\rm NS}} 
\approx 3 \times 10^{53} \,
\textrm{erg} \left( \frac{M_{\rm pns}}{M_\odot} \right)^2 
\left(\frac{r_{NS}}{12 \, \textrm{km}}\right)^{-1}.  
\end{equation} 
The \CCSN{} shock forms at an enclosed mass of $\sim 0.4 \, M_\odot$ and the
material that is shock heated increases the effective \PNS{} radius.  
This provides a reservoir of gravitational potential energy that can be
converted into neutrinos.  Therefore, around two thirds of the total energy,
$E_{\rm SN}$ is available during the \PNS{} cooling phase.

After the \CCSN{} shock has passed through the \PNS{}, the interior
entropy varies between one and six $k_b/\textrm{baryon}$.  Peak temperatures
between $30-60 \, \textrm{MeV}$ are reached during \PNS{} evolution, while
the surface of the \PNS{} has a temperature around $3-5 \, \textrm{MeV}$. 
The interior of the \PNS{} is comprised of
interacting protons, neutrons, and electrons, at densities greater than a
few times nuclear saturation density ($\rho_s \approx 2.8 \times 10^{14}
\textrm{g cm}^{-3}$) towards the center of the \PNS{}.  It is also possible that
more exotic degrees of freedom are present in the inner most regions of the
\PNS{} \citep{Prakash:97}.

%
%

\index{deleptonization}
During core collapse, electron capture on heavy nuclei removes around 40\% of
the electrons from the core before neutrinos become trapped \citep{Hix:03},
leaving behind $Y_e \approx 0.3$ in the core.  $Y_e$ is the number of electrons
per baryon and is equal to the proton fraction by charge conservation.  Although
this constitutes a large portion of the initial lepton number of the core, a
cold NS has an even lower total lepton number.  The lepton number of the \PNS{}
is the total number of electrons plus electron neutrinos minus the number of
positrons and electron antineutrinos, which is a conserved quantity.  In a cold
\NS{}, the interactions $e^- + p \rightarrow \nu_e + n$ and $n \rightarrow \bar
\nu_e + e^- + p$ are in equilibrium.  Equating these rates and solving for the
electron fraction results in $Y_e \sim 0.1$ for the densities encountered in the
cores of \NS{s}.  Therefore, the \PNS{} must``deleptonize'' to become a \NS{},
which requires losing a total lepton number of around 
\begin{equation}
N_L \approx 3.4 \times 10^{56} \left(\frac{M_{\rm pns}}{1.4 \, M_\odot} \right), 
\end{equation} 
which must be removed from the \PNS{} by neutrinos.  

\index{neutrino mean free path}
Inside the \PNS{}, a copious number of neutrinos of all flavors are produced and
scattered by weak interactions involving both the baryons and the leptons present in
the medium.  The rate at which neutrinos leave the \PNS{} and carry off energy
and lepton number will depend on thermal neutrino mean free path inside the
\PNS{} with energy $\epsilon_\nu \sim 60 \, \textrm{MeV}$.  Using a reference
weak interaction neutrino cross-section (see section \ref{sec:opacities})
\begin{equation}
\sigma_\nu = \frac{4 G_F^2 \epsilon_\nu^2}{\pi}
\approx 3 \times 10^{-40} \, \textrm{cm}^2 
\left(\frac{\epsilon_\nu}{60 \, \textrm{MeV}} \right)^2 \,,
\end{equation}  
where, $G_F$ is the Fermi coupling constant, $\epsilon_\nu$ is the neutrino
energy.  A naive estimate of the neutrino  mean free path
in the \PNS{} is then
\begin{equation}
\lambda_\nu \sim \frac{1}{\bar n_b \sigma_\nu} \approx 14 \, \textrm{cm} 
\left(\frac{\bar \epsilon_\nu}{60 \, \textrm{MeV}} \right)^{-2}
\left(\frac{R_{\rm pns}}{12 \, \textrm{km}} \right)^{3}
\left(\frac{M_{\rm pns}}{M_\odot} \right)^{-1},
\end{equation} 
where $\bar n_b = 3 M_{\rm pns}/(4\pi R_{\rm pns}^3 m_b)$ is the average baryon
density of the \PNS{} and $\bar \epsilon_\nu$ is a characteristic energy for
neutrinos inside the \PNS{}.  The neutrino mean free path is much
smaller than the radius of the \PNS{}, which is around 12 km once the shock
heated mantle has
cooled.  Therefore, neutrinos must escape from the \PNS{}
diffusively and will be in thermal and chemical equilibrium with the baryons and
electrons throughout most of the \PNS{}.  


\subsection{The Equations of \PNS{} Cooling} 
\label{sec:pns_cooling_eq}
\index{proto-neutron star cooling equations}
Generally, \PNS{} evolution is a neutrino radiation hydrodynamics problem where
general relativity is important.  A number of simplifications to the general
system of equations can be made.  First, the \PNS{} cooling time scale is much
longer than the sound crossing time of the \PNS{}.  Therefore, \PNS{s} are very
close to being in hydrostatic equilibrium and spherical symmetric.  With these
approximations, the equations of \PNS{} cooling become (see \citep{Burrows:86,
Pons:99, Roberts:12a} for detailed derivations)
\begin{eqnarray}
\label{eq:hydro1}
\frac{dP}{dr} &=& -\frac{G(M_g + 4\pi r^3 P)(\rho + P)}{r^2 \Gamma^2} \\
\label{eq:hydro2}
\frac{dM_g}{dr} &=& 4 \pi r^2 \rho \\
\label{eq:hydro3}
\frac{dN}{dr} &=& \frac{4 \pi r^2 n_b}{\Gamma} \\
\label{eq:hydro4}
\frac{d\alpha}{dP} &=& -\frac{\alpha}{P + \rho} \\
\label{eq:lepton}
\frac{dY_L}{dt} &=& -\frac{\partial (4 \pi \alpha r^2 F_L)}{\partial N}\\
\label{eq:ye}
\frac{dY_e}{dt} &=& \alpha \frac{S_N}{n_b} \\
\label{eq:energy}
\frac{d((\rho + \rho_\nu)/n_b)}{dt} &=& 
- \alpha^{-1} \frac{\partial (4 \pi \alpha^2 r^2 H_\nu)}{\partial N} 
- (P +  P_\nu) \frac{d(1/n)}{dt} \\
\label{eq:int_energy}
\frac{d(\rho/n_b)}{dt} &=& \alpha \frac{S_E}{n_b} - P \frac{d(1/n)}{dt}.
\end{eqnarray} 
Equations~\ref{eq:hydro1}, \ref{eq:hydro2}, \ref{eq:hydro3}, and \ref{eq:hydro4}
are just the relativistic equations of hydrostatic equilibrium, where $P$ is the
pressure, $\rho$ is energy density of the background fluid, $\rho_\nu$ is the
energy density of the neutrinos, $n_b$ is the baryon density,  $M_g$ is the
gravitational mass, $N$ is the enclosed baryon number, $r$ is the radius, $\Gamma =
\sqrt{1-2 G M_g/r}$ and $\alpha$ is the lapse function.  The lepton
fraction is $Y_L = Y_e + Y_{\nu_e}$, where $Y_e$ is the number of electron per
baryon and $Y_{\nu_e} = (n_{\nu_e} - n_{\bar \nu_e})/n_b$ is the local net number of electron neutrinos per baryon.
Equations~\ref{eq:lepton}, and \ref{eq:energy} describe conservation of lepton
number and total internal energy in the \PNS{}.  Throughout most of the \PNS{},
equations \ref{eq:ye} and \ref{eq:int_energy} are zero and can be neglected
since the neutrino number and energy source functions, $S_N$ and $S_E$, rapidly
bring the neutrinos into thermal equilibrium with the background fluid.  The
energy flux and lepton number fluxes are given by $F_L = F^N_{\nu_e} - F^N_{\bar
\nu_e}$ and $H_\nu = \sum F^E_i$ (where the sum runs over all flavors of
neutrinos and antineutrinos).  The number and energy fluxes of individual
neutrino species are given by
\begin{equation}
F_{\nu_i}^{\left\{N/E \right\}} = \frac{2 \pi}{(2 \pi)^3} 
\int_0^\infty d\epsilon \epsilon^{\left\{2/3 \right\}}
\int_{-1}^1 d\mu \mu f_{\nu_i},
\end{equation} 
where $f_{\nu_i} = f_{\nu_i}(t, r, \epsilon, \mu)$ is the distribution function
of neutrinos of species $i$, $\mu$ is the cosine of the angle of neutrino
propagation relative to the radial direction, and $\epsilon$ is the neutrino energy.
Below, we often discuss the neutrino luminosity, $L_{\nu_i} = 4 \pi r^2 \alpha^2
F^E_i$ and the neutrino number luminosity $\dot N_{\nu_i} = 4 \pi r^2 \alpha
F^N_i$.  

\index{diffusion approximation}
Solution of this system of equations requires a method of determining the
$f_{\nu_i}$, the evolution of which is determined by the Boltzmann
equation,
\begin{equation}
\frac{D f_{\nu_i}}{D l} = (\eta_{a,i} + \eta_{s,i}[f_{\nu_i}])  (1-f_{\nu_i}) 
- (\kappa_{a,i} + \kappa_{s,i}[f_{\nu_i}]) f_{\nu_i},
\end{equation}
where $D/Dl$ is a Lagrangian derivative in phase space, $eta_a$ and $\eta_s$ are
absorption scattering emissivities, and $\kappa_a$ and $\kappa_s$ are absorption and
scattering opacities \citep{Lindquist:66, Thorne:81}.  The evolution of the $f_\nu$ can be attacked directly
with the Boltzmann equation, but neutrino transport simplifies greatly throughout most of the
\PNS{}.  As was mentioned above, the neutrino mean free path inside the \PNS{}
is much shorter than the distance over which $n_b$, $T$, and $Y_e$ are
changing.  Therefore, the neutrino distribution functions are very
close to thermal and the neutrinos propagate through the star diffusively.  In
the diffusion limit diffusion limit of the Boltzmann equation, the number and energy flux of
neutrinos of species $i$ are given by opacity weighted radial derivatives of the
neutrino density \citep{Burrows:86, Pons:99}
\begin{equation}
F_{\nu i}^{\left\{N/E \right\}} = 
- \frac{\Gamma}{\alpha^{\{3/4\}}} \int^\infty_0 d\omega 
\frac{\omega^{\left\{2/3 \right\}}}
{3\left(\kappa^*_{a, i} + \kappa^*_{s, i}\right)} 
\frac{\partial f_{i,{\rm FD}}(\omega/\alpha)}{\partial r}, 
\end{equation}
where $f_{i,{\rm FD}}(\epsilon) = [1 + {\rm exp}(\epsilon/T - \eta_i)]^{-1}$ is
the Fermi-Dirac distribution for neutrinos of species $i$ with degeneracy
parameter $\eta_i$, $\omega$ is the neutrino energy at infinity.   Both electron
neutrinos and antineutrinos rapidly reach chemical equilibrium with the nuclear
medium via charged current neutrino interactions.  Therefore, the electron
neutrino chemical potential is $\mu_{\nu_e} = \mu_e + \mu_p - \mu_n$ and
$\mu_{\bar \nu_e} = - \mu_{\nu_e}$.  Because of the large mass of the $\mu$ and
$\tau$ particles, no net $\mu$ or $\tau$ number is produced in the \PNS{} and
$\mu_{\nu_\mu} = \mu_{\nu_\tau} = 0$.  The quantity $\kappa_{a,i}^*$ is the
total absorption opacity corrected for detailed balance and $\kappa_{s,i}^*$ is
the scattering transport opacity \citep{Pons:99, Burrows:06}. These opacities
have units of inverse length and are approximately $\kappa \sim n_b \sigma_\nu$.
In the equilibrium diffusion limit, the isotropic parts of the neutrino
distribution functions only depends on the local temperature, as well as the
electron neutrino degeneracy factor, $\eta_{\nu_e} = \mu_{\nu_e}/T$, for
electron neutrinos and antineutrinos.  The chemical potentials of the $\mu$ and
$\tau$ neutrinos are zero throughout the \PNS{} due to the large masses of the
$\mu$ and $\tau$ particles.  With these assumptions, the total lepton and energy
fluxes in the diffusion limit become
\begin{eqnarray}
F_L = 
- \frac{\Gamma T^2}{\alpha 6 \pi^2} \left[ 
D_3  \frac{\partial (\alpha T)}{\partial r} + 
D_2  \alpha T \frac{\partial \eta_{\nu_e}}{\partial r}
\right ] \\ 
H_\nu = - \frac{\Gamma T^3}{\alpha 6 \pi^2} \left[ 
D_4 \frac{\partial (\alpha T)}{\partial r} + 
D_3 \alpha T \frac{\partial \eta_{\nu_e}}{\partial r}
\right ], 
\end{eqnarray}
where $D_2 = D^{\nu_e}_2 + D_2^{\bar \nu_e}$, $D_3 = D^{\nu_e}_3 - D^{\bar
\nu_e}_3$, and $D_4 = \sum_i D^i_4$ are diffusion coefficients. 
The single species diffusion coefficients are
Rosseland mean opacities defined by 
\index{diffusion coefficient}
\begin{equation}
D^i_n = \int_0^\infty d\epsilon \frac{\epsilon^n f_{i,{\rm FD}}(1-f_{i,{\rm FD}})}
{T^{n+1}\left(\kappa^*_{a,i} + \kappa^*_{s,i}\right)}.
\end{equation} 
Only electron neutrinos and antineutrinos contribute to the lepton flux
diffusion coefficients while all species contribute to the energy flux diffusion
coefficients.  Note that the diffusion coefficients will be $\sim
\lambda_\nu(\langle \epsilon_\nu \rangle)$, where $\langle \epsilon_\nu \rangle$
is an average neutrino energy in the medium. This makes it clear that gradients
in the temperature and $\mu_{\nu_e}$ in the \PNS{} core, combined with the
neutrino opacities, drive its deleptonization and cooling. 

In addition to these structure and transport equations, a model for the dense
matter encountered in the \PNS{}, as well as the initial configuration of $s$
and $Y_L$ versus radius, is required to predict the \PNS{} neutrino signal.  The
\EOS{} -- which determines $\rho$, $P$, and $\mu_{\nu_e}$ as a function of
$n_b$, $T$, and $Y_e$ -- and the neutrino diffusion coefficients depend strongly
on the properties of matter at and above nuclear saturation density.  

The equilibrium diffusion equations described above provide an excellent
approximation to neutrino transport in the optically thick interior regions of
the \PNS{}, are useful for understanding the basic properties of \PNS{} cooling,
and have been used -- along with flux limiters to prevent superluminal transport
of energy and lepton number \citep{Burrows:86} -- in numerous works studying
the cooling of \PNS{s} \citep{Burrows:86, Keil:95, Pons:99, Roberts:12}.
Nonetheless, they are not suited to describing neutrino transport near the
surface where the neutrino mean free paths become large. They also cannot provide any
information about the average energies of the neutrinos that emerge from the
neutron star since they assume neutrinos are everywhere in thermal equilibrium
with the background material.  Therefore, some works have employed
non-equilibrium, spectral neutrino transport at varying levels of sophistication
\citep{Woosley:94, Sumiyoshi:95, Hudepohl:10, Fischer:10, Roberts:12b,
Nakazato:13}.  These methods all evolve the non-equilibrium distribution
function of the neutrinos, $f_{\nu_i}$, at a large number of neutrino energies
using the Boltzmann equation or some approximation thereof. 
 

\subsection{\PNS{} Evolution} 
\index{proto-neutron star evolution}
\label{sec:pns_evolution}
\begin{figure}
\includegraphics[width=\linewidth]{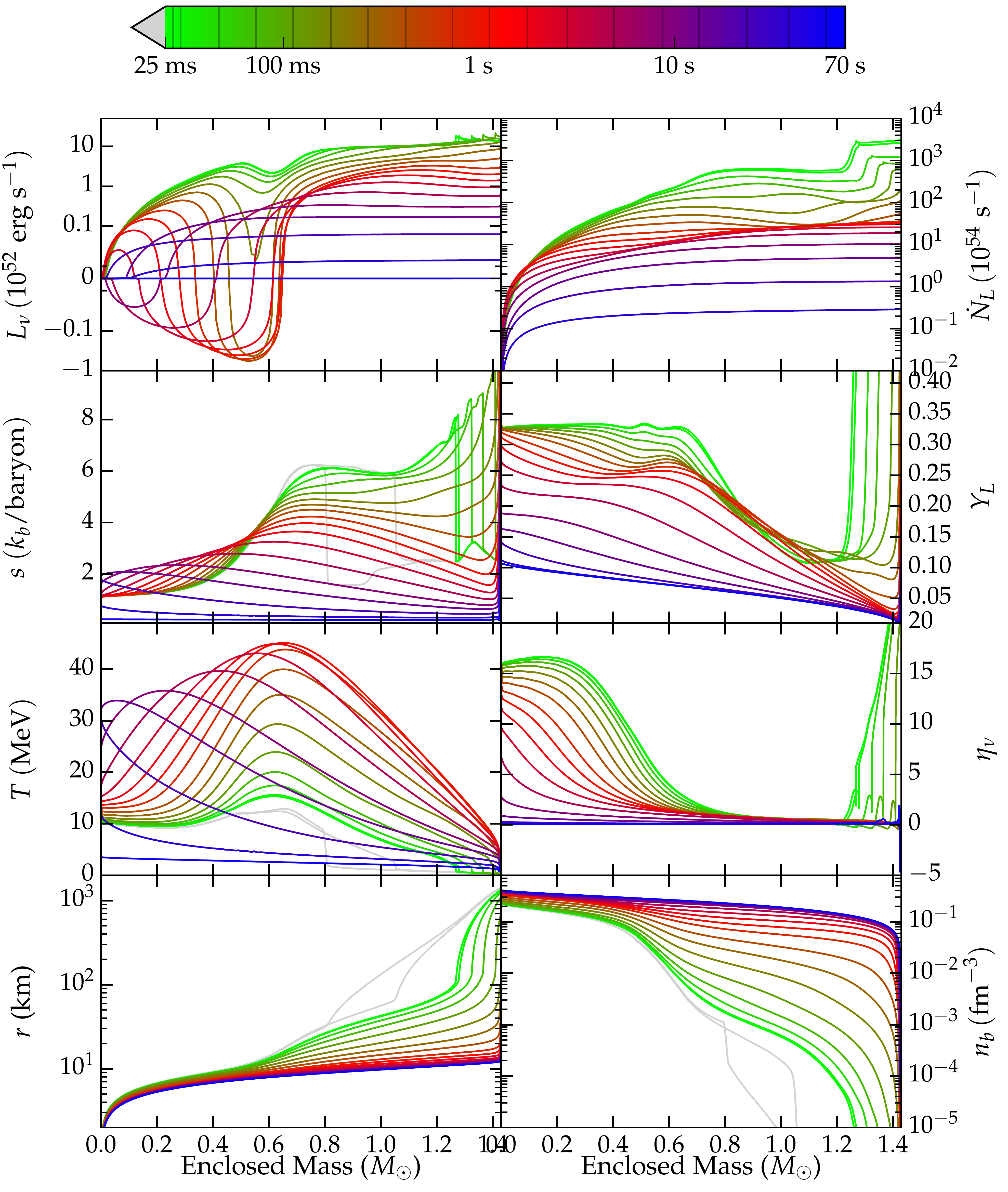}
\caption{Interior \PNS{} quantities at selected times, with time coded by color.
The semi-transparent lines on the color bar demarcate the time of the various
lines shown in the panels below it.  The gray lines are for selected times
during the dynamical post-bounce evolution.  Here, we focus on the evolution of
the inner core after the dynamical phase has ended.  The top left panel shows
the energy carried by all flavors of neutrinos while the top right panel shows
the net lepton number transported by neutrinos as a function of enclosed
baryonic mass.  The second row of panels shows the evolution of the entropy and
lepton fraction, which is determined by the neutrino fluxes.  The third row of
panels shows the temperature and neutrino degeneracy parameter evolution.
Gradients in these quantities drive the diffusive neutrino fluxes.  The final
row of panels shows the radius and baryon density as a function of enclosed
baryonic mass to illustrate how the structure of the \PNS{} evolves and contracts
in response to loss of energy and lepton number. } 
\label{fig:pns_interior}
\end{figure} 

\begin{figure}
\includegraphics[width=\linewidth]{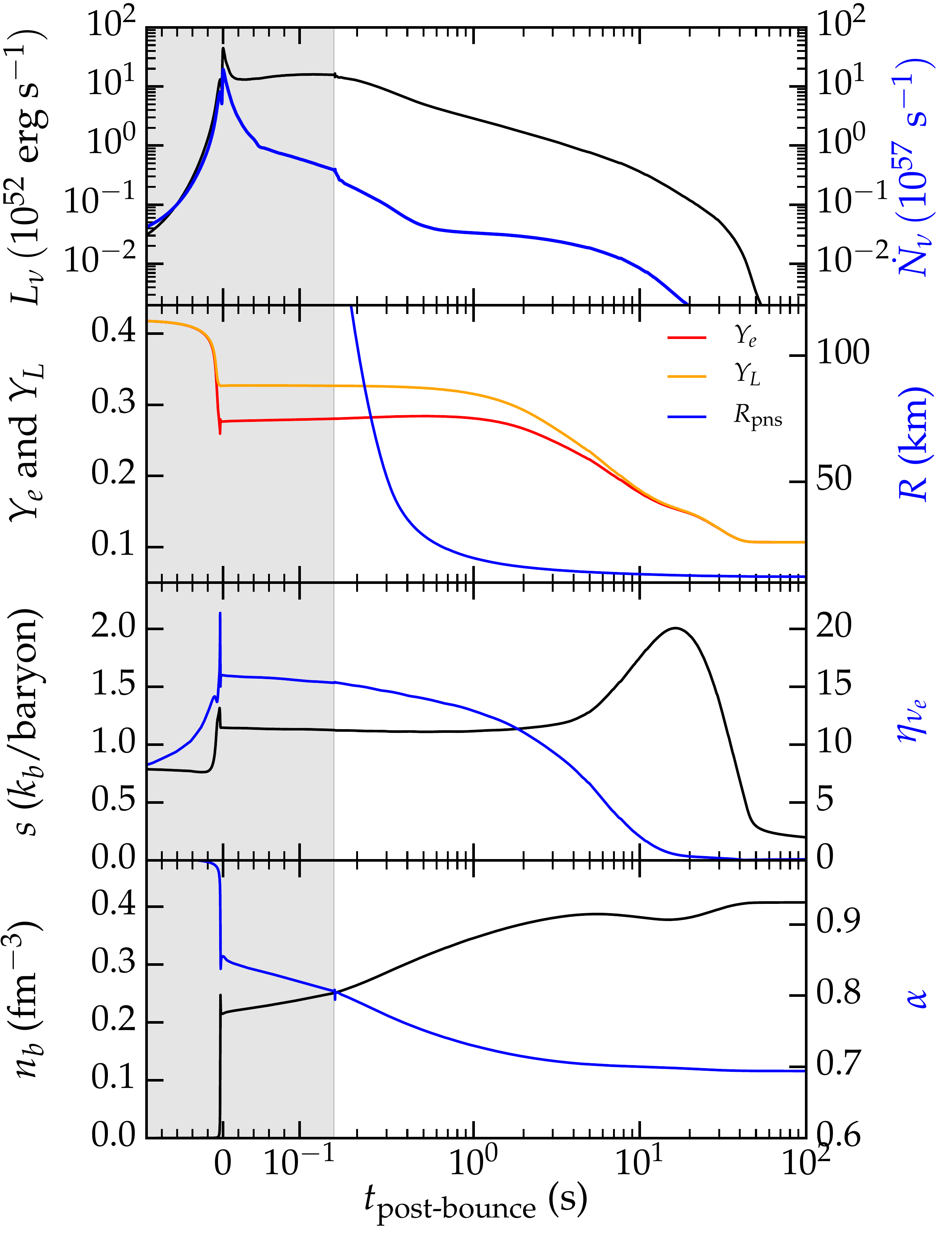}
\caption{Time evolution of central quantities and the total neutrino luminosity
and lepton flux.  The gray region corresponds to the accretion phase and is on a
linear time scale, while the region to the right is the \PNS{} cooling phase and
it is plotted on a logarithmic scale.  At the transition from the accretion
phase to the \PNS{} cooling phase, all of the material from above the shock is
excised from the grid, causing a slight jump in some quantities.  The top panel
shows the total energy loss rate from the \PNS{} and the deleptonization rate.
The second panel shows the evolution of the central lepton fraction and electron
fraction, as well as the \PNS{} radius.  The deleptonization era corresponds to
the period over which $Y_e$ and $Y_L$ differ.  The third panel shows the
evolution of the central neutrino chemical potential and entropy.  The impact of
Joule heating is visible between five and twenty seconds.  The bottom panel
shows the central density and the central lapse, $\alpha$, to illustrate the
contraction of the \PNS{} over time.} \label{fig:pns_time} \end{figure} 

Here, the evolution of a fiducial $1.42 \, M_\odot$ \PNS{} model is described.
\PNS{} evolution has been modeled using numerical codes for almost thirty years
\citep{Burrows:86, Burrows:87, Sumiyoshi:95, Keil:95, Keil:95a, Pons:99, Pons:01,
Pons:01, Fischer:10, Hudepohl:10, Roberts:12, Roberts:12a, Nakazato:13}.
Although the fidelity to the underlying micro- and macro-physics has improved
with time, the basic features of \PNS{} cooling are still the same.  The
illustrative model discussed in this section was calculated using the
multi-energy group neutrino transport method described in \citep{Roberts:12b}.
Even so, throughout most of the \PNS{} this more realistic transport methods
reduces to the diffusion equations described above.

This calculation began with a pre-\SN{} model of a $15 \, M_\odot$  star from
\citep{Woosley:95}.  The evolution through core collapse, bounce, and the
cessation of shock expansion was followed.  Once the \SN{} shock crossed an
enclosed baryonic mass of $1.42 \, M_\odot$, material outside this region was
excised from the grid to crudely simulate the \CCSN{} explosion.  Then this \PNS{}
core was evolved for one hundred seconds. This model is comparable to models found
in \citep{Hudepohl:10, Fischer:10}.  We used an \EOS{} similar to the
Lattimer-Swesty equation of state with incompressibility $K_s=220 \,
\textrm{MeV}$ \citep{Lattimer:91} and all neutrino-nucleon interactions were
treated in the elastic limit.  The evolution of the interior structure of the
model is shown in figure \ref{fig:pns_interior}.  The time evolution of various
quantities in the center of the \PNS{} as well as its energy and lepton number
losses are shown in figure \ref{fig:pns_time}.  These figures also show the
evolution of the star during the pre-explosion phase of the SN (see the chapter
``Neutrino emission from supernovae'' for a detailed discussion of the early
time emission).

The initial structure left behind after deleptonization during core collapse and
shock propagation through the core can be seen with the yellow lines in the
panels showing the entropy and the lepton fraction in figure
\ref{fig:pns_interior}.  There is a low entropy, high lepton number {\it core}
and a high entropy, low lepton number {\it mantle}.  The transition point
between these two phases is near the maximum temperature point in the \PNS{}.
After collapse and bounce, the \SN{} shock formed near this transition point, so
the mantle is shock heated and the core is not.  The high entropy found in the
mantle increases its pressure and thereby increases its radius and decreases 
its density relative to the core.  

The initial core lepton fraction, $Y_L = 0.33$, is set by the point during core
collapse when neutrinos become trapped, which occurs when the core reaches a
density near $10^{-4} \, {\rm fm}^{-3}$ \citep{Hix:03}.  This can be
seen in the second and fourth panels of figure \ref{fig:pns_time}.  Electron
neutrinos are able to rapidly remove lepton number from the mantle, so this
region has $\eta_{\nu_e} \approx 0$.

Once the \SN{} shock begins to explode the star and accretion slows down or
ceases, \PNS{} neutrino emission proceeds in three phases: the mantle
contraction phase, the deleptonization phase, and the thermal cooling phase.
The basic features of these different phases can be seen in figures
\ref{fig:pns_interior} and \ref{fig:pns_time}.

\index{proto-neutron star mantle}
{\it Mantle Contraction} - During the first few seconds, the neutrino emission
is dominated by contributions from the contracting mantle.  This contraction is
not dynamical.  Rather, it is driven by the reduction in pressure due to entropy
and electron losses. As can be seen in the second to last panel of figure
\ref{fig:pns_time}, the radius of the \PNS{} contracts from around 100 km to a
value very close to the cold NS radius -- which is 12 km for the \EOS{} state used
in this model -- within the first two seconds of \PNS{} evolution.  The neutrino
luminosity emerging from the surface of the \PNS{} is sourced completely in this
region, see the top left panel of figure \ref{fig:pns_interior}.  In fact,
energy is also being lost from the inner boundary of the mantle as electron
antineutrinos and heavy flavored neutrinos diffuse down the positive temperature
gradient into the core; the total neutrino luminosity becomes negative at the
core mantle boundary. Additionally, the deleptonization wave pushes inward over
this period but does not reach the center of the \PNS{}.  This can be seen in the
panels of figure \ref{fig:pns_interior} showing $Y_L$ and $\eta_{\nu_e}$.  

{\it Deleptonization Phase} - Once the \PNS{} has settled down to close to the
radius of a cold NS, the luminosity evolution is driven by the deleptonization
wave that propagates into the center of the \PNS{} and eventually brings the entire
\PNS{} to a state were $\eta_{\nu_e}=0$.  This occurs over a period of about twenty
seconds.  The lepton number flux is produced by the negative gradient in
$\eta_{\nu_e}$ left after shock breakout in the region between the homologous
core and the base of the mantle.

\index{Joule heating}
During this period the core entropy and temperature increase, as can be seen in
the bottom panel of figure \ref{fig:pns_time}.  This is due to
two effects.  First, as was the case during the mantle contraction phase,
electron antineutrinos and heavy flavored neutrinos are diffusing inward and
heating the core.  Second, lepton number is being lost from the core due to the
positive flux of electron neutrinos throughout the \PNS{}, which causes ``Joule
heating'' \citep{Burrows:86}.  To see this, we recast equation \ref{eq:energy}
in terms of the entropy using the first law of thermodynamics
\begin{equation}
\label{eq:entropy}
T  \frac{ds}{dt} = 
- \alpha^{-1} \frac{\partial (4 \pi \alpha^2 r^2 H_\nu)}{\partial N} 
- \mu_{\nu_e} \frac{d Y_L}{dt}.
\end{equation}
Here, $s$ is the entropy per baryon in units of Boltzmann's constant, $T$ is the
temperature, and $\mu_{\nu_e}$ is the electron neutrino chemical potential.
Joule heating comes from the second term on the right hand side of this
equation, since $dY_L/dt<0$ and $\mu_{\nu_e}>0$.  Eventually, the combined
effects of these two processes raise the central entropy from around one to two
$k_b/\textrm{baryon}$ and create a negative entropy and temperature gradients
throughout the star.  

{\it Thermal Cooling Phase} - Once $\eta_{\nu_e} \sim 0$ throughout the star
(see figure \ref{fig:pns_interior}), the \PNS{} slowly contracts as energy leaks
from the entire star.  Both the entropy and lepton number of the core fall
during this period, as shown in figure \ref{fig:pns_time}.  The deleptonization
rate falls off rapidly, but it does not go to zero because the local equilibrium
electron fraction decreases with the local temperature, so low level
deleptonization continues.  The period of \PNS{} cooling ends when the object
becomes optically thin to neutrinos and the neutrino luminosity falls off
abruptly.

\subsection{Analytic Estimates of Cooling Phase Timescales}
\label{sec:timescales}
\index{neutrino cooling timescale}
It is instructive to use the equations of section \ref{sec:pns_cooling_eq} and a
few simplifying assumptions to obtain analytic solutions to the neutrino transport
equations.  This can elucidate how the timescales of the deleptonization and
thermal cooling phases depend on the neutrino opacities and the \EOS{}.  

At the onset of deleptonization electron neutrinos are degenerate.  Under these
conditions, the gradient in neutrino chemical potential dominates lepton number
flux in Eq.~\ref{eq:lepton}. Further, due to Pauli blocking only neutrinos at
the Fermi surface contribute and the relevant diffusion coefficient reduces to  
\begin{equation} 
D_2\simeq \frac{\mu^2_{\nu_e}}{2T^2}~\frac{1}{\kappa_a^*(\mu_{\nu_e})}\,.
\label{eq:d2}
\end{equation} 
In degenerate matter, as we shall show later in section \ref{sec:opacities}   
\begin{equation} 
\kappa_a^* (\mu_{\nu_e},T) \approx \frac{G^2_F}{2\pi} 
~\frac{1+3g^2_A}{4}~M^2T^2\mu_e \simeq \left( \frac{k_B T}{15 ~{\rm MeV}}\right)^2  
\left(\frac{\mu_e}{200 ~{\rm MeV}} \right)~\frac{1}{20~{\rm cm}}
\end{equation}  
where $M$ is the nucleon mass, $T$ is the temperature and $\mu_e$ is the
electron chemical potential. Using Eq.~\ref{eq:d2} and neglecting general
relativistic corrections we, can approximate the lepton number flux in
Eq.~\ref{eq:lepton} as
\begin{equation} 
F_{L} \approx \frac{n_b}{6\kappa^*_a}~\frac{\partial Y_{\nu_e}}{\partial r}\,.
\end{equation} 
By noting that $n_b/\kappa^*_a$ is a weak function of the density we ignore its
spatial dependence to find an analytic solution of the separable form
$Y_\nu(r,t)=Y_{\nu,0}\phi(t)\psi(r)$ to Eq.~\ref{eq:lepton}, similar to the
method descibed in \citep{Prakash:97}.  Using appropriate
boundary conditions at the surface we separately solve for spatial and temporal
dependencies with $\phi(0)=1$ and $\psi(0)=1$.  For the temporal part, which is
of interest here, we obtain a simple exponential solution 
\begin{eqnarray}
\phi(t) = \exp \left({\frac{-t}{\tau_D}}\right) \quad {\rm where} \quad
\tau_D \simeq  \frac{6 \langle \kappa^*_a \rangle ~R^2}{C_x} 
\frac{\partial Y_L}{\partial Y_\nu} \,.
\label{dlepdt}
\end{eqnarray}
Here, $ \langle \kappa^*_a \rangle$ represents a spatial average of the charged
current opacity inside the \PNS{} and the constant $C_x\simeq 10$ depends on
$\psi(r)$.  Using fiducial values $T=15$ MeV  and $\mu_e=200$ MeV and $\partial
Y_L/\partial Y_\nu = 5$ and setting $\langle \kappa^*_a \rangle =
\kappa^*_a(\mu_e=200~{\rm MeV},T=15 ~{\rm MeV}) $ we obtain 
\begin{equation}
\tau_D \approx 11  \left(\frac{R}{10~{\rm km}}\right)^2
\left(\frac{ k_BT}{15~{\rm MeV}}\right)^2 \left(\frac{\mu_e}{200~{\rm MeV}}\right)  
\left(\frac{\partial Y_L}{ 5~\partial Y_\nu}\right) \quad s \,.
\end{equation}
This result, albeit arrived at with some approximation, clearly reveals the
microphysics. The dependence on $T$, $\mu_e$, and $\partial Y_L/\partial Y_\nu$
is made explicit and we discuss later in section \ref{sec:eos} how the dense
matter \EOS{} directly affects these properties.  

We can also estimate the amount of Joule heating in the core
(see equation \ref{eq:entropy})
\begin{eqnarray}
\dot{E}_{joule} = - n_b \mu_\nu \frac{\partial Y_L}{\partial t} 
\approx  n_b \mu_\nu \frac{\partial Y_L}{\partial Y_\nu}\frac{Y_{\nu,0}}{\tau_D} \,,
\end{eqnarray}
where we have used Eq.~\ref{dlepdt} to express the result in terms of the
deleptonization time.  For typical values of the deleptonization time $\tau_D
\sim 11$ s, and $\partial Y_L/\partial Y_\nu \sim 5$, we find the heating rate
per baryon $\dot{E}_{joule}/n_b \approx \mu_\nu Y_{\nu,0}/3 $.  At early times
when $\mu_\nu \sim 150$ MeV and $Y_{\nu,0}\sim 0.05$ the heating rate $\approx
2$ MeV per baryon per second will result in a similar rate of change in the
matter temperature. This, coupled with the positive temperature gradients,
results in a net heating of the inner core when $t<\tau_D$.  

After deleptonization when the core begins to cool, the second term in
Eq.~\ref{eq:entropy} can be neglected and the energy flux 
\begin{equation}
H_\nu \approx \frac{T^3}{6\pi^2}~D_4
\frac{\partial T}{\partial r} \,.
\label{eflux}
\end{equation}
Energy transport is dominated by $\nu_\mu,
\bar{\nu}_\mu,\nu_\tau,\bar{\nu}_\tau$ and $\bar{\nu}_e$ neutrinos since their
charged current reactions are suppressed and therefore they have larger mean
free paths.  For typical conditions where nucleons are degenerate and neutrino
degeneracy is negligible, elastic neutral current scattering off nucleons is
dominant source of opacity and (see section \ref{sec:opacities})   
\begin{equation}
\kappa^*_{s}(E_\nu)\simeq \frac{5}{6\pi}~ G^2_F~c^2_A~\tilde{N}_0 ~k_BT~E^2_\nu \,, 
\label{kappa_s}
\end{equation}
where $ \tilde{N}_0  = \sum_{i=n,p} \partial n_i/\partial \mu_i$ is the
effective density of nucleon states at the fermi surface to which neutrinos
couple, and $c_{A}\simeq 1.2$ is the axial vector coupling. Using
Eq.~\ref{kappa_s} the diffusion coefficient $D_4$  in Eq. \ref{eflux} can be
written as 
\begin{equation}
D_4=\frac{\pi^3}{G^2_Fc^2_A~\tilde{N}_0~(k_BT)^3}\,.  
\label{eq:d4} 
\end{equation}
Substituting Eq.~\ref{eq:d4} in Eq.~\ref{eflux}, Eq.~\ref{eq:entropy} can be
solved with the separable ansatz $T(r,t)=T_c \psi(x) \phi(t)$ to find a
self-similar solution. We find that the temporal part  $\phi(t)=1-(t/\tau_c)$,
where   
\begin{equation}
\tau_c \approx \frac{2 \pi G^2_F c^2_A}{ \beta} \left\langle N_0
\frac{3 n_b}{\pi^2}\frac{\partial s}{\partial T}\right \rangle
~k_BT_c ~R^2 \simeq  10 ~s~\frac{k_BT_c}{30~{\rm MeV}}~
\frac{ \langle n^{2/3} _b \rangle}{n_0^{2/3}}~\left(\frac{R}{12~{\rm km}}\right)^2  
\,,
\end{equation}
where $\langle   \rangle$ denotes a spatial average, the numerical constant
$\beta \cong 19$, and $n_0=0.16$ fm$^{-3}$.  Additionally, we have used
$\partial s/\partial T = \pi^2 N_0/3 n_b$ and $N_0 = M(3 \pi^2n_b)^{1/3}/\pi^2$,
which hold for a non-relativistic, degenerate gas. The spatial averages and
numerical value of $\beta$ are obtained by solving for the function $\psi(r)$.  

\subsection{\PNS{} Neutrino Emission}
\label{sec:pns_neutrino_emission}
\begin{figure}
\includegraphics[width=\linewidth]{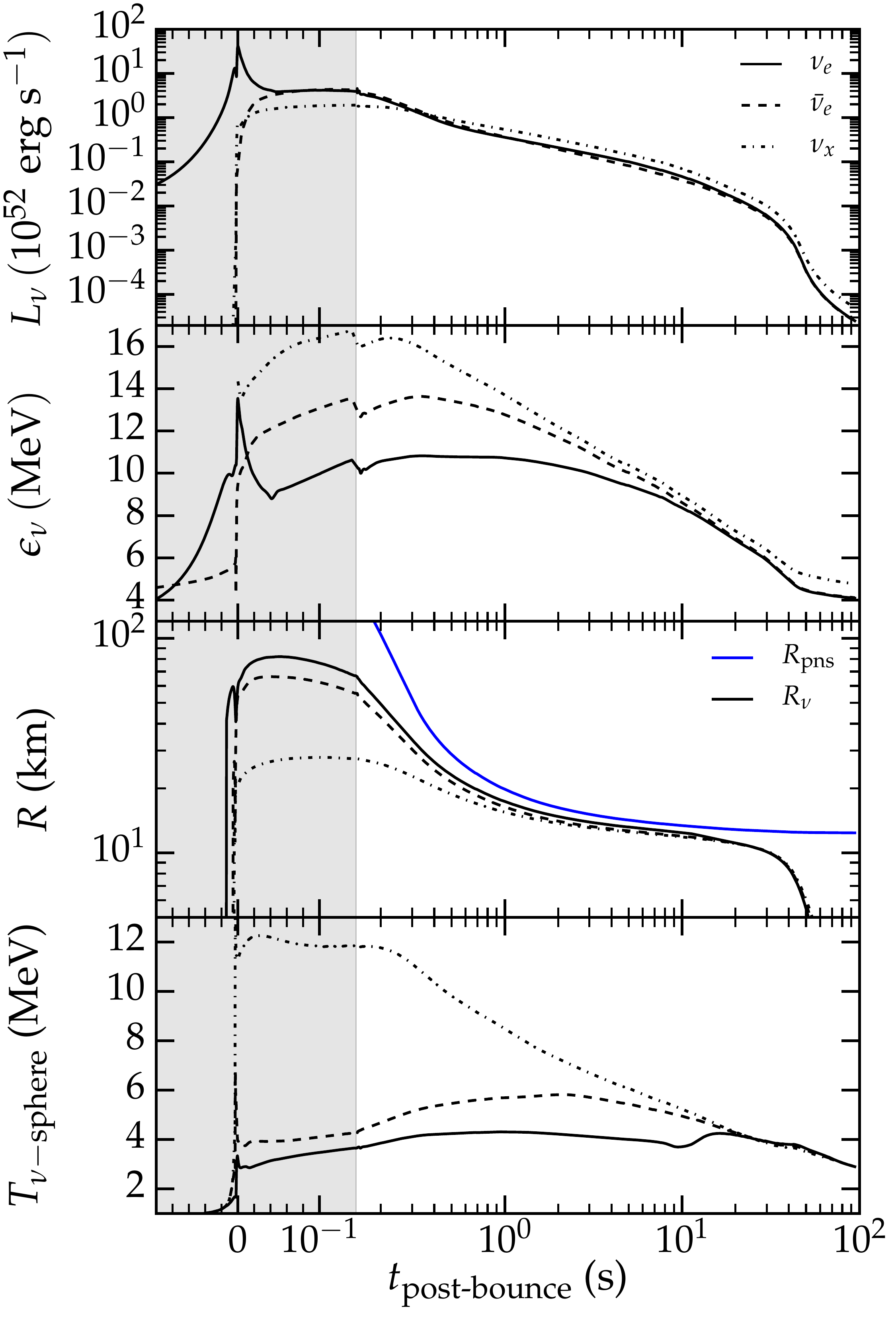}
\caption{Time evolution of neutrino properties at infinity and the properties of
the neutrino spheres.  The gray region corresponds to the accretion phase and is on a
linear scale, while the region to the right is the \PNS{} cooling phase and it is
plotted on a logarithmic scale.  We measure radiation quantities at a radius of
300 km or at the maximum radius of the simulation domain, whichever is smaller.
At the transition from the accretion phase to the \PNS{} cooling phase, all of the
material from above the shock is excised from the grid, causing a slight jump in
some quantities.  In all of the panels, all heavy flavored neutrinos are
represented by $\nu_x$, since their properties are all very similar.  In the top
panel, we show the neutrino luminosity for single flavors as a function of time.
In the second panel, we show the evolution of the neutrino average energies.  In
the third panel we show the evolution of the neutrino sphere radii, along with
the \PNS{} radius, as a function of time.  The simulation covers the entire time
the \PNS{} is optically thick to neutrinos.  The bottom panel show the
temperature at the neutrinospheres. Similar results can be found in
\citep{Fischer:11}.} \label{fig:pns_nusphere} \end{figure} 


Here, we discuss the evolution of the flavor dependent neutrino luminosities and
average energies, which constitute the detectable signal from the \PNS{}.  We focus
on the properties of the neutrinos near the surface of the \PNS{}.  As the
neutrinos propagate through the rest of the star, neutrino oscillations can
change the flavor content of the neutrino fields and alter the spectra of the
neutrinos that eventually reach earth \citep{Duan:10}.  The effects of
neutrino oscillations on the \PNS{} signal are discussed in the chapter
``Neutrino conversion in supernovae''.  

\index{neutrino sphere} 
The evolution of the interior of the \PNS{} drives the total energy and lepton
number emitted from the surface, but the outermost layers of the \PNS{} shape the
spectrum of the emitted neutrinos.  Therefore, it is convenient to describe the
approximate radius at which neutrinos that escape to infinity are emitted, the
``neutrino sphere'' $R_\nu$.  Following \citep{Keil:03, Fischer:11}, we define
\begin{equation} 
\tau_{\textrm{therm}}(R_\nu) = \int_{R_\nu}^\infty dr
\sqrt{\langle \kappa_a \rangle ( \langle \kappa_a \rangle + \langle \kappa_s
\rangle)} = \frac{2}{3}, 	
\end{equation} 
where $\langle \kappa \rangle$ is an opacity averaged over the local neutrino
distribution function.  This is of course an approximation, since neutrino
interactions have a strong energy dependence.  Neutrinos of the same flavor but
different energies will therefore decouple at different positions within the
\PNS{}.  Nonetheless, the neutrino luminosity in a particular flavor can be
reasonably estimated by assuming the neutrino sphere is a black-body emitter
\begin{equation}
L_\nu = 4 \pi \sigma_{BB} \phi R_\nu^2 T(R_\nu)^4,
\end{equation}
where $\sigma_{BB} = 4.75 \times 10^{35} \textrm{erg MeV}^{-4} \textrm{ cm}^{-2}
\textrm{ s}^{-1}$ is the black body constant and $\phi$ is a factor of order
unity that accounts for deviations from strict Fermi-Dirac black body emission
\citep{Hudepohl:10, Mirizzi:15}.  For a pure black body spectrum, the average
energy of the emitted neutrinos would be $\langle \epsilon_\nu \rangle \approx
3.15 T(R_\nu)$.  
The energy moments of neutrinos of species $i$ at radius $r$ are given by
\begin{equation}
\langle \epsilon_{\nu_i}^n \rangle = 
\frac{\int_0^\infty d\epsilon \int_{-1}^1 d\mu \epsilon^{n+2} f_{\nu_i}}
{\int_0^\infty d\epsilon \int_{-1}^1 d\mu\epsilon^2 f_{\nu_i}}.
\end{equation}
In reality, high energy neutrinos have a larger
decoupling radius than lower energy neutrinos due to the approximate
$\epsilon_\nu^2$ scaling of the neutrino opacities \citep{Keil:03}.  Therefore,
high energy neutrinos are emitted from regions with lower temperatures.  The
emitted neutrino spectra then have a ``pinched'' character, where there is a
deficit of high energy neutrinos relative to the Fermi-Dirac spectrum predicted
by an energy averaged neutrinosphere \citep{Janka:89}.  

\index{neutrino average energy}
The evolution of the neutrino sphere radii as a function of time are shown in
the third panel and the temperature at the neutrino spheres is shown in the
fourth panel of figure \ref{fig:pns_nusphere}.  There are only small differences
between the $\mu$ and $\tau$ flavored neutrino and antineutrino emission because
they experience similar neutral current opacities.  Therefore, we group all of
these flavors together in flavor $x$.  During the accretion phase and
into the mantle contraction phase, there is a clear hierarchy with $R_{\nu_x} <
R_{\bar \nu_e} < R_{\nu_e}$.  This is driven mainly by differences in the
charged current opacities: electron neutrinos get a large opacity contribution
from the reaction $\nu_e + n \rightarrow e^{-} + p$ due to the large number of
neutrons present near the decoupling region, electron antineutrinos get a
somewhat smaller contribution from $\bar \nu_e + p \rightarrow e^+ + n$ because
of the small number of protons present in the decoupling region, and heavy
flavored neutrinos receive no contribution to their opacity from charged
current interactions.  Since the temperature is decreasing with increasing
radius, this gives rise to the standard early time hierarchy of neutrino
energies $\langle \epsilon_{\nu_e} \rangle < \langle \epsilon_{\bar \nu_e}
\rangle < \langle \epsilon_{\nu_x} \rangle$ during the accretion and mantle
cooling phases that can be seen in the second panel of figure
\ref{fig:pns_nusphere}.  During this period, inelastic scattering from electrons
outside of the neutrino sphere reduces the average energies of
heavy flavored neutrinos relative to what the black body model would predict
\citep{Raffelt:01}.  In the deleptonization phase, there are very few protons in the
outer layers of the \PNS{}.  Therefore, the
opacities for the electron antineutrinos and the heavy flavored neutrinos are
very similar and all of these neutrino flavors decouple at similar radii.  In
figure \ref{fig:pns_nusphere}, these neutrinos have almost equal
average energies over the majority of the \PNS{} cooling phase.  
Eventually, during the thermal cooling phase all three neutrino spheres converge
and the average emitted energies of all flavors become similar
\citep{Fischer:11}, although the time at which they converge depends on how the
charged current opacities are calculated \citep{Roberts:12}.  

\index{neutrino luminosity}
The top panel of \ref{fig:pns_nusphere} shows the luminosities of individual
flavors of neutrinos.  Soon after shock break out, the electron neutrino and
antineutrino luminosities become very close to one another and stay similar
throughout the entire cooling evolution.  Deleptonization proceeds because the
average energies of the electron neutrinos are lower and more electron neutrinos
are required to carry a fixed luminosity than electron antineutrinos.  Due to
their small neutrinosphere, the heavy flavored neutrino luminosities are much
lower than the electron neutrino luminosities during the mantle cooling phase.
After mantle contraction, there is approximate equipartition of luminosity
among the different flavors. 

\section{Physics that shapes the cooling signal}
\label{sec:eos}
\index{nuclear equation of state}
\index{nuclear symmetry energy}
The \PNS{} neutrino signal is interesting in part because it is shaped by the
properties of neutron rich material at densities and temperatures that
are inaccessible in the laboratory. Because of the high densities encountered
throughout the \PNS{}, the inter-nucleon separation is small enough that
interactions between nuclei play a central role in determining the \PNS{}
\EOS{} and the neutrino interaction rates. Over the past decade
improved models to describe hot and dense matter were developed that reproduce
empirically known properties of symmetric nuclear matter at saturation density.
However, since matter encountered in the proto-neutron stars is characterized by a 
small proton fraction $x_p \simeq 0.05-0.3$, the symmetry energy defined through
the relation 
\begin{equation} 
S(n_b) = E(n_b,x_p=1/2) - E_n(n_b,x_p=0)\,,
\label{eq:esym}
\end{equation} 
where $E(n_b,x_p=1/2)$ is the energy per particle of symmetric nuclear matter
and $E(n_b,x_p=0)$ is the energy per particle of pure neutron matter, plays an
important role. The energy of neutron-rich matter $E(n_b,x_p) \simeq
E(n_b,x_p=1/2) + S(n_b)(1-2x^2_p)$ since quartic and higher order terms are
found to be small in most theoretical calculations.  In this context,  {\it ab
intio} calculations of the energy of neutron matter at sub-saturation density
\citep{Gandolfi:11,Tews:2012fj} have provided valuable guidance in the
development of a new suite of models for hot and dense matter which are also
consistent with recent neutron star radii in the range $10-12$ km. These smaller
radii are favored by recent modeling efforts to extract the radius of neutron
stars from x-ray observations of of quiescent neutron star in low mass x-ray
binaries and x-ray burst \citep{Steiner:13,Ozel:2015fia}. Properties of matter at
and around nuclear saturation density, especially $S(n_b)$, can influence
deleptonization and neutrino cooling timescales, convection and the neutrino
spectrum \citep{Sumiyoshi:95, Roberts:12, Roberts:12b}. PNS evolution is also
sensitive to the thermal properties of degenerate dense matter as discussed in
\ref{sec:timescales}, and for a discussion of it we refer the reader to
Refs.~\citep{Prakash:97, Constantinou:2014hha, Rrapaj:2015zba}. In the next two
subsections, we describe how nuclear interactions can affect neutrino opacities
and convective instabilities inside the \PNS{}, both of which can alter the
\PNS{} cooling timescale.  

\index{hyperons}
\index{quark matter}
At densities near and below nuclear saturation density, \PNS{} matter is only
composed of protons, neutrons, and electrons, but at higher densities it is
possible for more exotic material to be present.  Hyperons--baryons containing
strange quarks--can be produced in the interior of the \PNS{} because the weak
interaction does not conserve strangeness \citep{Prakash:97}.  Quark matter
\citep{Steiner:01, Pons:01a} or Bose condensates \citep{Prakash:97, Pons:00a,
Pons:01} may also exist in the inner most regions of \PNS{s}.  In addition to
altering the neutrino opacities, these new degrees of freedom serve to
soften the nuclear \EOS{} at high density and reduce the maximum
neutron star mass.  This can lead to ``metastable'' \PNS{s} that emit neutrinos
for tens of seconds before collapsing to a \BH{} when more exotic material forms
in their core and pressure support is reduced.  \BH{} formation would abruptly
end the neutrino signal and is therefore directly observable \citep{Pons:01a}.
We do not discuss these effects in any more detail here, but refer the reader to
\citep{Prakash:97}.     

\subsection{Neutrino opacities in dense matter}
\label{sec:opacities}

\begin{figure}
\includegraphics[width=\linewidth]{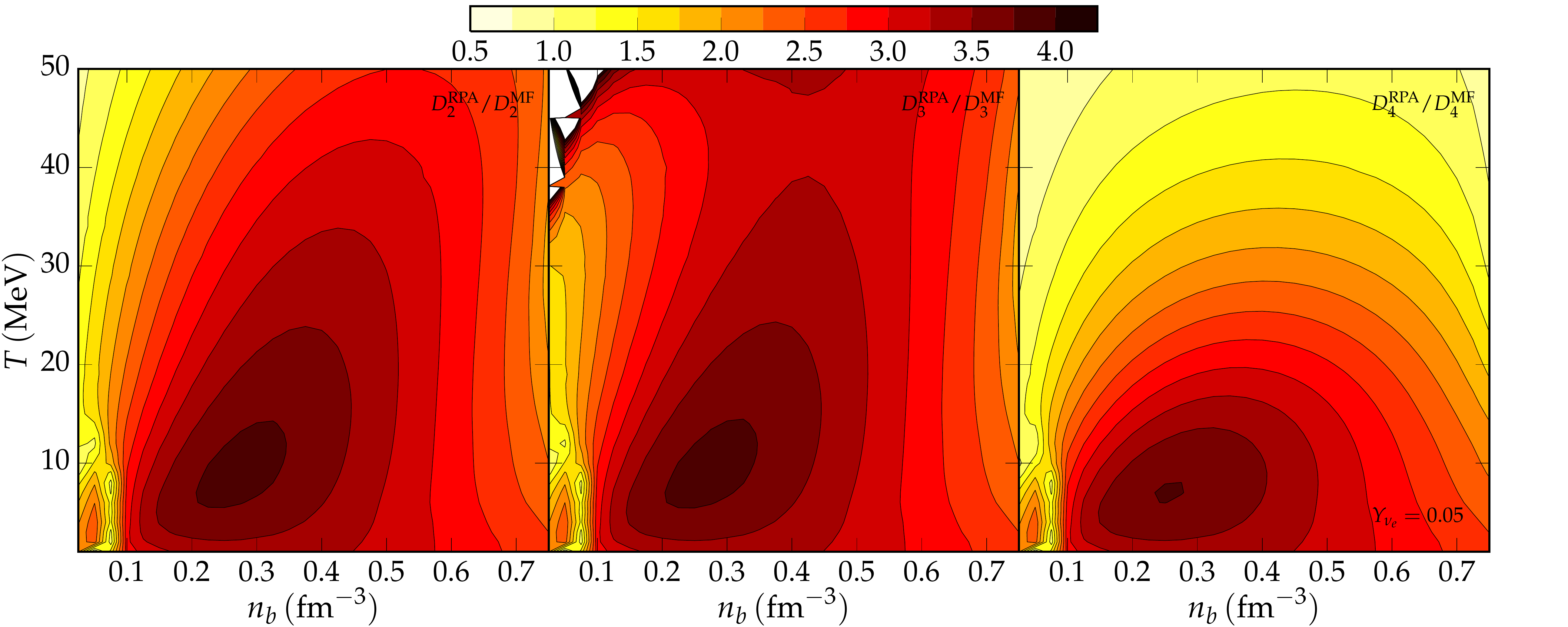}
\caption{Ratio of diffusion coefficients not including nuclear correlations to
diffusion coefficients including nuclear correlations.  At high density,
weak charge screening (calculated using the random phase approximation)
supresses the neutrino opacity and increases the neutrino diffusion
coefficients.  This reduces the \PNS{} cooling timescale.}
\label{fig:diffusion_coefs}
\end{figure}

\begin{figure}
\includegraphics[width=\linewidth]{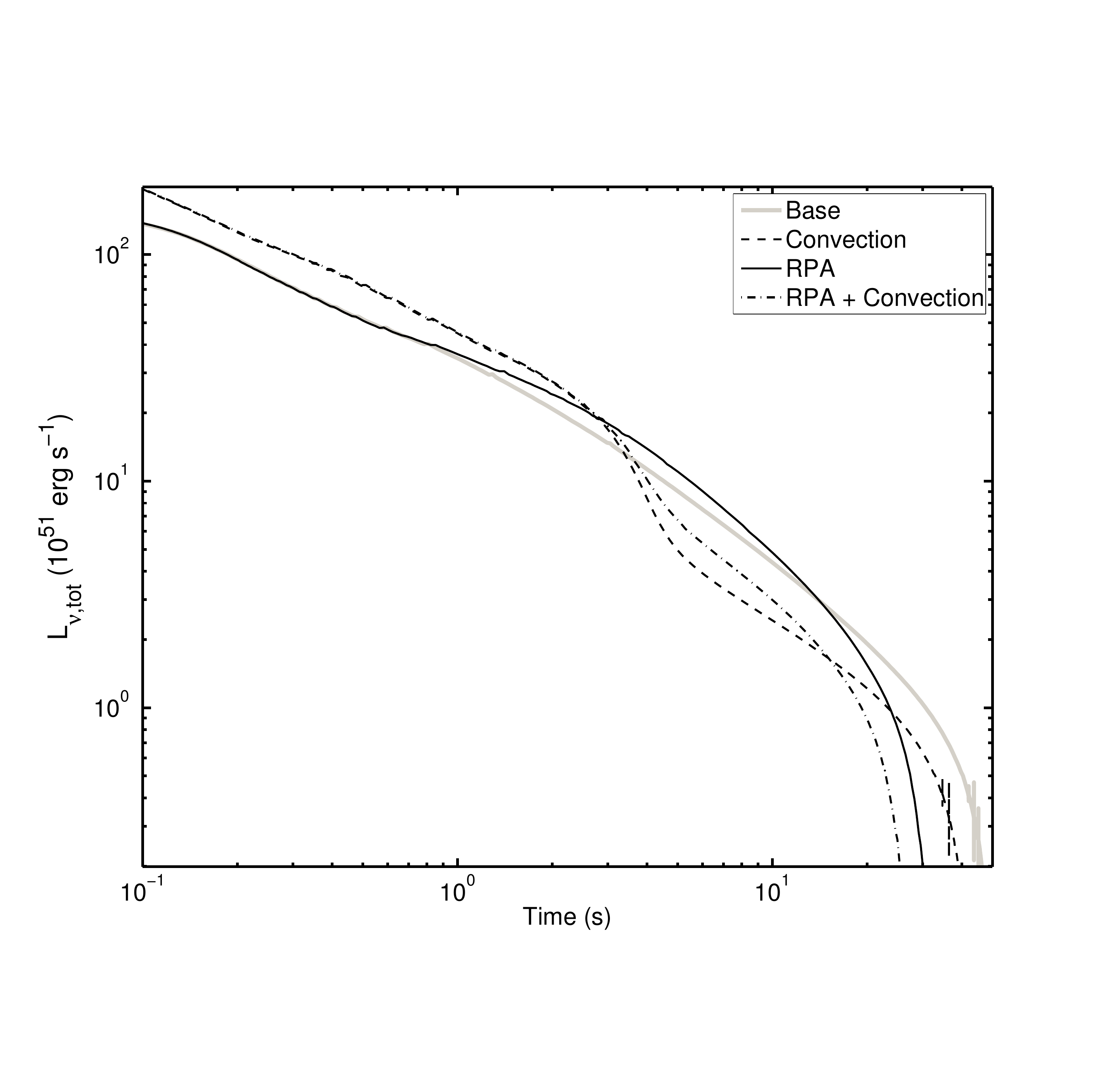}
\caption{The total \PNS{} neutrino luminosity versus time for a number of \PNS{}
models that include convection and/or the affect of nuclear correlations on the
opacity.  Both convection and nuclear correlations decrease the cooling
timescale relative to the baseline model.  Convection alters the luminosity at
early times, while correlations only become important after the mantle cooling
phase.  The models shown here are similar to those described in
\citep{Roberts:12}.}
\label{fig:luminosities}
\end{figure}

\index{neutrino interactions}
The neutrino scattering opactities and thereby the diffusion coefficients
defined in section \ref{sec:pns_cooling_eq} receive contributions from neutrino
scattering, absorption, and pair-production processes, as well as the inverses
of the latter two.  Scattering contributions come from the reactions
\begin{eqnarray*}
\nu_i + n &\rightleftharpoons&   \nu_i + n \\
\nu_i + p &\rightleftharpoons&   \nu_i + p \\
\nu_i + e^- &\rightleftharpoons& \nu_i + e^- \\  
\nu_i + e^+ &\rightleftharpoons& \nu_i + e^+, 
\end{eqnarray*}
as well as scattering from other possible components of the medium.  All of the
reactions above have neutral current contributions for all flavors of neutrinos,
while $e^-$ and $e^+$ scattering also have a charged current contribution for 
$\nu_e$ and $\bar \nu_e$ scattering, respectively.  Since the dominant
scattering contribution for all particles comes from the $n$ and $p$ scattering,
there are only small differences between the scattering contributions to the
diffusion coefficients for different neutrino flavors.  Scattering from
electrons and positrons can be highly inelastic, due to the small mass of the
electron relative to the characteristic \PNS{} neutrino energy, while scattering
from neutrons and protons is close to elastic.  This inelasticity can
alter the emitted neutrino spectrum and serves to bring
the average energies of the different neutrino species closer to one another
\citep{Hudepohl:10}.  

The diffusion coefficients for the various neutrino flavors become 
different from one another due to charged current neutrino interactions.  The
main absorption contribution to $D_i^{\nu_e}$ comes from
\begin{equation}
e^{-} + p \rightleftharpoons \nu_e + n, \nonumber
\end{equation}
while the main absorption contribution to $D_i^{\bar \nu_e}$ comes from 
\begin{equation}
e^{+} + n \rightleftharpoons \bar \nu_e + p. \nonumber
\end{equation}
All of the opacities receive contributions from thermal processes
such as 
\begin{eqnarray}
N + N &\rightarrow& N + N + \nu + \bar \nu \nonumber \\
e^- + e^+ &\rightarrow& \nu + \bar \nu, \nonumber
\end{eqnarray}
but these are usually small compared to the charged current interactions that
affect the electron neutrinos and antineutrinos.  

\index{neutrino opacity}
For both scattering and absorption processes, the cross section per unit volume
for a general process $\nu + 2 \rightarrow 3 + 4$ (where particle 3 is either a
neutrino, electron, or positron) can be written as \citep{Reddy:98}
\begin{eqnarray}
\kappa(\epsilon_\nu) &=& \frac{2}{2 \epsilon_\nu (2\pi)^9} 
\int \frac{d^3 p_2}{2 \epsilon_2 } 
\int \frac{d^3 p_3}{2 \epsilon_3 }  
\int \frac{d^3 p_4}{2 \epsilon_4 }
f_2(\epsilon_2)  (1-f_3(\epsilon_3)) (1-f_4(\epsilon_4)) 
\nonumber\\ &&\times 
(2 \pi)^4 \delta^4 \left (P_\nu + P_2 - P_3 - P_4 \right ) 
\langle \left | \mathcal{M} \right |^2 \rangle, 
\end{eqnarray}
where 
\begin{eqnarray}
\langle \left | \mathcal{M} \right |^2 \rangle &=&
16 G_F^2 \bigl[ 
(C_V^2+C_A)^2 (P_\nu \cdot P_2)(P_3 \cdot P_4) 
+(C_V-C_A)^2 (P_2 \cdot P_3)(P_\nu \cdot P_4) \\
&&-(C_V^2-C_A^2) M_2 M_4 (P_\nu \cdot P_3)
\bigr] 
\end{eqnarray}
is the spin summed weak interaction matrix element of the process, $C_V$ and
$C_A$ are vector and axial coupling constants, $\delta^4$ is the
four-dimensional Dirac delta function, $P_i = (\epsilon_i, -\vec{p}_i)$ is the
relativistic four-momentum, $p_i$ is the three-momentum, $\epsilon_i$ is the
energy, and $f_i$ is the distribution of species $i$.  Particles 2 and 4 are
always in thermal equilibrium inside \PNS{s}, so $f_2$ and $f_4$ are isotropic
Fermi-Dirac distribution functions.    This expression comes from directly from
Fermi's Golden Rule.  If we specialize to particle 2 and 4 being nucleons, the
reduced matrix element becomes independent of the nucleon momenta at leading
order $v/c$ and the cross-section can be written as 
\begin{equation}
\kappa(\epsilon_\nu) = \frac{G_F^2}{(2\pi)^2}  
\int_{-1}^1 d \mu_3  (C_V^2(1+\mu_3) + C_A^2(3-\mu_3))
\int_0^\infty d\epsilon_3 p_3 \epsilon_3  
(1-f_3(\epsilon_3))
S(q_0, q),
\end{equation} 
where $C_V$ and $C_A$ are weak vector and axial-vector coupling constants of the
weak interaction, $\mu_3 = \vec{p_\nu} \cdot \vec{p_3}/(|\vec{p_\nu}|
|\vec{p_3}|)$, and  
\index{response function}
\index{structure function}
\begin{equation}
S(q_0, q) = 2 
\int \frac{d^3 p_2}{(2 \pi)^3} 
\int \frac{d^3 p_4}{(2 \pi)^3}
f_2(\epsilon_2) (1-f_4(\epsilon_4)) 
(2 \pi)^4 \delta^4 \left (Q + P_2 - P_4 \right ) 
\end{equation}
The energy-momentum transfer from the neutrino to the nucleons is denoted by
the four-vector $Q=(q_0,-\vec{q})$, such that $\epsilon_3 = \epsilon_\nu - q_0$
and $q = |\vec{q}| = \sqrt{\epsilon_\nu^2 + \epsilon_3^2 - 2\epsilon_\nu
\epsilon_3 \mu_3}.$  This
form of the opacity separates the contribution of the nucleons (or the
``medium'') from the neutrino and the outgoing particle (be it another particle
or a neutrino with a different energy).  The function $S(q_0,q)$ is often
referred to as the response function or structure factor.  A similar separation
is found when the full momentum dependence of the matrix element is included,
although there are multiple response functions with different kinematic
dependence \citep{Reddy:98}.  The contribution to the cross section from particle
3 is the amount of phase space available to it in the final state, which results
in the leading order $\epsilon_\nu^2$ dependence of weak interaction cross
sections when $q_0\sim0$.

\index{response function, non-interacting}
The response function includes the effects of energy/momentum conservation,
Pauli blocking of the final state nucleons, and thermal motion of the nucleons.
When the momentum transfer $q$ is small (so that $\vec{p}_2 = \vec{p}_4$), the
response simplifies significantly.  After manipulating the Fermi-Dirac
distribution functions and integrating, one finds 
\begin{equation}
S(q_0, 0) = 2 \pi \delta(q_0) \frac{n_2 - n_4}{1-\exp((\mu_4 - \mu_2)/T)}.
\end{equation}
When species 2 and 4 are non-degenerate, this reduces to $2\pi \delta(q_0) n_2$.
This response is purely elastic, since $q_0=0$ and $\epsilon_\nu=\epsilon_3$. 
For scattering reactions, when species two equals species four, this response
becomes 
\begin{equation}
S_{\rm scat}(q_0, 0) = 
2 \pi \delta(q_0) T \frac{\partial n_2}{\partial \mu_2}.
\end{equation}
In the degenerate limit, the response can be shown to be \citep{Reddy:98}
\begin{equation}
S_{\rm deg}(q_0,q) = \frac{M^2 (q_0 + \mu_2 - \mu_4)}{\pi q} 
\frac{\Theta(q - p_{F_2} + p_{F_4})}{1 - \exp((\mu_4 - \mu_2 - q_0)/T)},
\end{equation}
where $p_{F_i}$ is the Fermi momentum of particles of species $i$ and $\Theta$
is the Heaviside step function.  The opacity then becomes 
\begin{equation}
\kappa(\epsilon_\nu) = \frac{G_F^2}{4 \pi^3}(C_V^2 + 3 C_A^2)
M^2 T^2 (\epsilon_\nu + \mu_2 - \mu_4) \Xi
\frac{\pi^2 + \left(\frac{\epsilon_\nu - \mu_\nu}{T} \right)^2}{1+\exp((\mu_\nu -
\epsilon_\nu)/T)},
\end{equation}
where 
\begin{equation}
\Xi= \Theta(p_{F_4}+p_{F_3} - p_{F_2} - p_{F_\nu}) + \frac{p_{F_4} + p_{F_3} -
p_{F_2} + p_{F_\nu}}{2 \epsilon_\nu} 
\Theta(p_{F_\nu} - |p_{F_4} + p_{F_3} - p_{F_2}|).
\end{equation}
When neutrinos are degenerate, only neutrinos near the Fermi surface will be
able to scatter.  The relevant opacity is then 
\begin{equation}
\kappa(\mu_\nu) = \frac{G_F^2}{8 \pi} (C_V^2 + 3C_A^2) M^2 (k_BT)^2 \mu_3,
\end{equation} 
where we have assumed all four species are in equilibrium, $\mu_\nu + \mu_2 =
\mu_3 + \mu_4$.  These results are used in section \ref{sec:timescales} to
estimate the deleptonization and thermal cooling timescales.

\index{response function, mean field approximation}
\index{mean field approximation}
If nucleon-nucleon interactions are also considered, they can alter the response
of the medium in a number of ways \citep{Hannestad:98, Reddy:98,
Reddy:99, Burrows:98, Horowitz:03}.  The simplest way to include
nucleon-nucleon interactions is in the mean field approximation.  In this
approximation, the averaged interaction with all other nucleons gives single
nucleons a momentum independent potential energy and an effective in medium
mass.  The nucleon energy-momentum relation then becomes $\epsilon_{2,4} =
p^2_{2,4}/2m^*_{2,4} + U_{2,4}$.    In the zero momentum transfer limit, the
response becomes \citep{Reddy:98}
\begin{equation}
S_{\rm MF}(q_0, 0) = 2 \pi \delta(q_0 + \Delta U) 
\frac{n_2 - n_4}{1-\exp((\mu_4 - \mu_2 + \Delta U )/T)},
\end{equation}
where $\Delta U = U_2 - U_4$.
This implies that $\epsilon_3 = \epsilon_\nu + \Delta U$.  Because the
cross-section strongly depends on the phase space available to particle 3, a
large, positive $\Delta U$ can increase the neutrino cross
section while a negative $\Delta U$ will reduce the cross section
\citep{Martinez-Pinedo:12, Roberts:12b}.  The potential energy of neutrons,
$U_n$, differs from the potential energy of protons, $U_p$ due to the isospin
dependence of the nuclear interaction.  In neutron rich material, neutrons have
a larger potential energy than protons because of the large, positive nuclear
symmetry energy, $S(n_b)$, throughout the \PNS{}.  In fact, the potential energy
difference can be related directly to the nuclear symmetry energy
\citep{Hempel:15}.  Therefore, mean field corrections to the response in \PNS{s}
increase the cross section for $\nu_e + n \rightarrow e^- + p$ and decrease the
cross section for $\bar \nu_e + p \rightarrow e^+ + n$.  This change alters
$D_2$ and $D_3$ and thereby the \PNS{} deleptonization rate \citep{Roberts:12a}.
These corrections can also move the electron neutrino sphere to a larger radius
and the electron antineutrinosphere to a smaller radius.  This increases the
difference between $\langle \epsilon_{\nu_e}\rangle$ and $\langle \epsilon_{\bar
\nu_e}\rangle$, which may have large consequences for nucleosynthesis near the
\PNS{} (see section \ref{sec:nucsyn}) \citep{Roberts:12a, Martinez-Pinedo:12}. 

\index{nucleon correlations}
\index{response function, random phase approximation}
When the neutrino wavelength is long compared to the inter-nucleon separation
distance, neutrino interactions with the medium concurrently involve multiple
nucleons at a microscopic level.  In this limit, collective properties induced
by nuclear interactions can significantly alter the response of the nuclear
medium.  The mean field approximation does not account for possible
nucleon-nucleon correlations induced by interactions.  Generally, accounting for
these correlations is a complex many-body problem which has only been tackled
within the random phase approximation (RPA) \citep{Burrows:98,
Reddy:99}.  The RPA essentially accounts for weak charge screening, which can
reduce the neutrino opacity.  In figure \ref{fig:diffusion_coefs}, we show the
suppression of the diffusion coefficients by correlations.  At high density, the
corrections can be larger than a factor of two.  This serves to decrease the
cooling timescale of the \PNS{} \citep{Burrows:98, Reddy:99, Hudepohl:10,
Roberts:12}.  In figure \ref{fig:luminosities}, we show models of \PNS{} cooling
that include RPA corrections to the neutrino interaction rates compared with
those that do not.  During the mantle contraction phase they have little affect
because the neutrino luminosity originates in a low density region.  Once the
neutrino luminosity is determined at higher densities, during the
deleptonization and thermal cooling phases, these corrections decrease the
neutrino emission timescale.  The magnitude of this effect depends on the
assumed nucleon-nucleon interaction \citep{Keil:95a, Reddy:99, Roberts:12}.

\subsection{\PNS{} Convection} 
\label{sec:pns_convection}

\index{proto-neutron star convection}
In addition to neutrinos, hydrodynamic motions of the \PNS{} can transport
energy and lepton number through the star.  Although the majority of the \PNS{}
is in hydrostatic equilibrium, there can be regions which are unstable to the
development of convection.  Similar to the case in normal stellar burning,
convective overturn can transport energy and lepton number much more rapidly
than radiation and shorten the cooling timescale \citep{Burrows:87, Wilson:88,
Roberts:12}.  

The standard Ledoux criterion for convective instability, adapted to the
conditions found inside a \PNS{}, is given by 
\begin{equation}
\omega_C^2 = -\frac{g}{\gamma_{n_b}}\left(\gamma_s \nabla\ln(s)
+ \gamma_{Y_L} \nabla \ln(Y_L) \right) > 0,
\end{equation}
where $g$ is the local acceleration due to gravity and 
\begin{eqnarray*}
\gamma_{n_b} = \thd{\ln P}{\ln n_b}{s,Y_L},
\gamma_{s} = \thd{\ln P}{\ln s}{n_b,Y_L},
\gamma_{Y_L} = \thd{\ln P}{\ln Y_L}{n_b,s}.
\end{eqnarray*}
These last three quantities are only functions of the nuclear \EOS{}.  In
particular, $\gamma_{n_b}$ is related to the sound speed and $\gamma_{n_b}$ and
$\gamma_{s}$ are always positive.  The third thermodynamic derivative,
$\gamma_{Y_L}$ can either be positive or negative; the pressure receives
contributions from both electrons which have $\partial P_e / \partial Y_e>0$ and
from the nucleons for which $\partial P_N / \partial Y_e <0$ when $Y_e<0.5$ due
variations in $\partial S(n_b)/\partial n_b$ where $S(n_b)$ is the nuclear symmetry energy defined in Eq.~\ref{eq:esym}  \citep{Roberts:12}. Noting that $\partial Y_e /
\partial Y_L > 0$ and of order unity, it is easy to see that the sign of
$\gamma_{Y_L}$ can change depending on the relative contributions of electrons
and nucleons.  Therefore, the portion of the \PNS{} that is convectively unstable
depends on the assumed nuclear \EOS{} and its symmetry energy, as well as the
gradients of entropy and $Y_L$
\citep{Roberts:12}.  The \PNS{} may also be subject to doubly diffusive
instabilities due to the lateral transport of composition and energy by
neutrinos \citep{Wilson:88}.  These double
diffusive instabilities would extend the region over which the \PNS{} was unstable
to hydrodynamic overturn. 

The outer \PNS{} mantle is
unstable to adiabatic convection soon after the passage of the bounce shock
\citep{Epstein:79}.  This early period of instability beneath the neutrinospheres
has been studied extensively in both one and two dimensions with the hope that
it could increase the neutrino luminosities enough to lead to a successful
\CCSN{} explosion \citep{Wilson:88, Buras:06}.  During the deleptonization and
thermal cooling phases, more and more of the \PNS{} becomes unstable to
convection because of the entropy and lepton gradients produced by neutrino
cooling.  Figure \ref{fig:pns_interior} shows that there are negative entropy
gradients throughout the mantle for the entirety of the \PNS{} phase and the
negative gradient extends through the whole \PNS{} by the end of
deleptonization.   

Because the cooling timescale is much longer than the dynamical timescale of the
\PNS{}, multi-dimensional simulations of late-time \PNS{} convection have not been
carried out to date.  Rather, mixing length theory has been employed to study
the impact of convection on the late time neutrino signal \citep{Burrows:87,
Roberts:12, Mirizzi:15}.  In figure \ref{fig:luminosities}, we show the total
neutrino luminosity for models with and without convection to highlight the
impact convection can have on the cooling timescale.  At early times, the
luminosity is elevated by around $25\%$ due to convection until the period of
convective instability ceases a few seconds after bounce.  After this the
luminosity is depressed relative to the case without convection, and the overall
cooling timescale is reduced. 

\section{Observable Consequences}
\subsection{Neutrinos from \SN{} 1987A}
\label{sec:1987a_neutrinos}

\begin{figure}
\includegraphics[width=\linewidth]{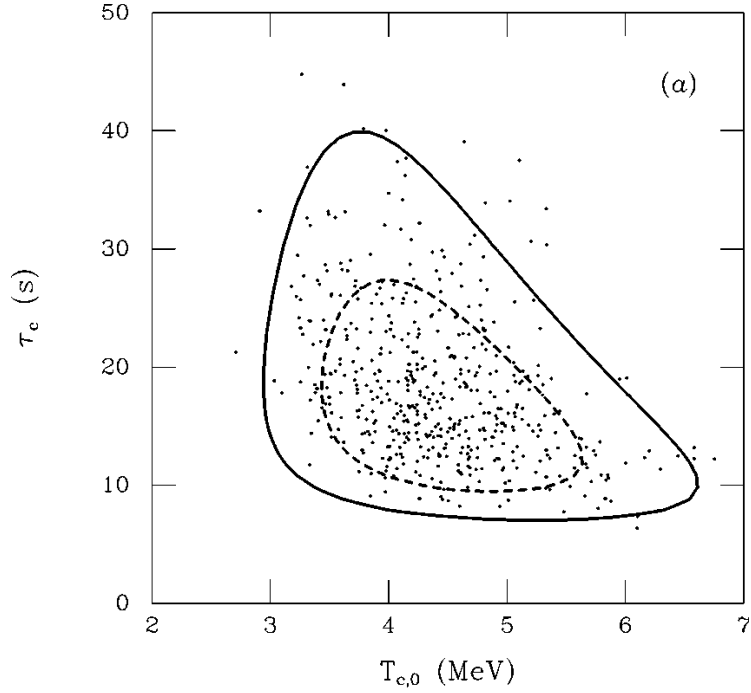}
\caption{Probability contours of the \PNS{} cooling timescale, $\tau_c$, and the
anti-electron neutrino spectral temperature $T_{c,0}$ base on the \SN{} 1987a
neutrinos observed at the Super Kamiokande, Baksan, and IMB detectors, taken
from \citep{Loredo:02}.  The dashed and solid lines demarcate regions of 68\% and
95\% credibility for the \PNS{} cooling parameters, respectively.  These
inferred parameters are in good qualitative agreement with the predictions from
models of \PNS{} cooling.  In this model, the average electron antineutrino
energy is given by $\langle \epsilon_{\bar \nu_e} \rangle \approx 3.15 T_{\bar
\nu_e} \approx 13.7 \, \textrm{MeV}$, where the last value is the value at one
second post-bounce.  Reprinted figure with permission from Loredo \& Lamb, Phys.
Rev. D, Vol. 65, 063002, (2002).  Copyright 2002 by the American Physical
Society.} \label{fig:1987Params} \end{figure} 

\index{proto-neutron star cooling}
\index{neutrinos, SN 1987A}
Up to now, our discussion of \PNS{} cooling has been mostly theoretical.  In fact,
the basics features of \PNS{} cooling were reasonably well characterized before
there was any observational evidence for this picture \citep{Burrows:86}.  In
February of 1987, both photons \citep{Kunkel:87} and neutrinos \citep{Bionta:87,
Hirata:87, Alexeyev:88} reached Earth from a massive star that collapsed in the
Large Magellanic Cloud, \SN{} 1987A.  Of course, the neutrinos are of the
most interest for constraining \PNS{} evolution, since the optical depth near the
\PNS{} is far too high for photons to escape this region.  Bursts of sixteen, eight,
and five neutrinos were observed near the time of \SN{} 1987A in the Kamiokande
\citep{Hirata:87}, IMB \citep{Bionta:87}, and Baksan \citep{Alexeyev:88} neutrino
detectors, respectively.  Although there were only $\sim 30$ electron
antineutrino events detected, they provided general confirmation of the
previously developed theoretical picture of neutrino emission during core
collapse and subsequent \PNS{} cooling \citep{Burrows:87a}.  The neutrinos were
observed within a 23 second window at Kamiokande and had energies ranging
between 5 and 30 MeV.  It is worth mentioning that this implies a \PNS{} survived
in the core of \SN{} 1987A for at least twenty seconds, even though no central object
has been observed in the remnant of \SN{} 1987A to date \citep{Graves:05}.  

In figure \ref{fig:1987Params}, the joint probability distribution of
the electron antineutrino spectral temperature and \PNS{} cooling timescale
inferred from the detected \SN{} 1987A neutrino events by \citep{Loredo:02}.  They
model the \PNS{} cooling phase with a luminosity falling off with time as $L_{\bar
\nu_e}(t) \propto (1 + t/\tau_c)^{-4}$ and Fermi-Dirac neutrino spectrum with
temperature $T_{\bar \nu_e} = T_{c,0}(1+t/\tau_c)^{-1}$.  The neutrino
cooling timescale
is best fit with $\tau_c = 14.7 \, \textrm{s}$ and the antineutrino average
energy at one second after bounce is best fit by $\langle \epsilon_{\bar \nu_e}
= 13.7 \, \textrm{MeV}$.    This is in reasonable qualitative agreement with the
models in the literature and the one presented in section
\ref{sec:pns_neutrino_emission}.  It is quite an achievement that models of \PNS{}
cooling based on theoretical considerations alone before 1987 were able to
reproduce the general features of the \SN{} 1987A neutrino signal.  Never the less,
there is uncertainty in the cooling timescale and $T_{c,0}$.
Additionally, attempts have been made to constrain various physical processes
operating during \PNS{} cooling using the \SN{} 1987A neutrino data
\citep{Burrows:87a, Keil:95a, Reddy:99}.

\index{axions}
The neutrino cooling signal from \SN{} 1987A has also been used to constrain beyond
the standard model physics.  Essentially, it is possible for exotic particles
with very weak -- but not too weak -- interactions to rapidly remove energy from
the \PNS{} \citep{Raffelt:96}.  If the amount of energy removed is comparable to the
amount of energy emitted in neutrinos, the neutrino cooling timescale can be
shortened.  This technique has been mainly used to put limits on the properties
of axions using the \SN{} 1987A neutrino signal \citep{Keil:97}. 

\subsection{Galactic Supernova Neutrinos}
\label{sec:galactic_sn_neutrinos}
Any future galactic \SN{} will yield far more neutrino detections than \SN{}
1987A.  
The expected rate of
\CCSN{} in the Milky Way is around $1-2$ per century \citep{Cappellaro:01}, so direct
\CCSN{} detection is somewhat of a waiting game.  Because of this, there are no
dedicated \CCSN{} neutrino detectors, but luckily there are many neutrino
experiments that can moonlight as \CCSN{} neutrino observatories. 

\index{neutrino detection rate}
The neutrino detection rate for a galactic \CCSN{} can be found by
integrating the neutrino distribution function at Earth over the response of the
neutrino detector \citep{Pons:99, Scholberg:12},  
\begin{eqnarray}
\td{N}{t} &=& \frac{2 \pi n_d}{(2 \pi)^3} \int_0^\infty d \epsilon_\nu
\epsilon_\nu^2 \sigma_\nu(\epsilon_\nu) W(\epsilon_\nu) \int_{-1}^1 d\mu
f_\nu(\epsilon_\nu, \mu, D)
\nonumber\\
&\approx& 87.5 \, {\rm s}^{-1} 
\left(\frac{D}{10 \textrm{kpc}}\right)^{-2} 
\left(\frac{\mathcal{M}_{\rm det}}{32 \, \textrm{kt}} \right)
\left(\frac{L_{\bar \nu_e}}{10^{51} \textrm{erg s}^{-1}}\right) 
\left(\frac{\langle \epsilon_{\bar \nu_e} \rangle}{12 \textrm{MeV}}\right)
G[f_{\bar \nu_e}(\epsilon)],
\end{eqnarray}
where $D$ is the distance from Earth to the \SN{}, $n_d$ is the number of
particles available to interact with neutrinos in the detector, and
$W(\epsilon)$ is the detector efficiency as a function of energy.  In the second line, we estimate the
electron antineutrino detection rate in a Cerenkov water detector with detector
mass $\mathcal{M}$.  The detector parameters chosen are meant to approximately
correspond to the properties of Super-Kamiokande \citep{Fukuda:03}.  The
dimensionless factor 
\begin{equation}
G[f_{ \nu}(\epsilon)] =
\frac{m_e^2 \int_0^\infty d \epsilon \epsilon^2 
\sigma_\nu(\epsilon) W(\epsilon) f_{\nu}}{ 
\sigma_\nu(\epsilon = m_e)  \langle \epsilon_\nu \rangle^2 
\int_0^\infty d \epsilon \epsilon^2 f_{\nu}}
\end{equation}
encodes the spectral distribution of the neutrinos folded with the neutrino
cross section and detector response.  The presence of a detector threshold in
$W(\epsilon)$ can make the dependence of the count rate on the neutrino average
energy steeper than is suggested by the simple scaling relation above.  For a
Fermi-Dirac distribution of neutrinos with zero chemical potential, $W$ has a
value $\approx 1.8$.  Integrating over a predicted \CCSN{} neutrino signal,
\citep{Scholberg:12} find that around 7,000 electron antineutrino events would be
detected in Super-Kamiokande for a \SN{} 10 kpc away and that the majority of
these neutrinos would come from the \PNS{} cooling phase \citep{Mirizzi:15}.
Additionally, electron neutrinos will be detectable in liquid argon detectors through
the reaction $\nu_e+^{40}Ar\rightarrow e^- + ^{40}K$.  For a supernova at 10 kpc,
we can expect about $700$ events per kiloton \citep{Scholberg:12}. The $\simeq
40$ kiloton liquid Argon detector planned at the Deep Underground Neutrino
Experiment (DUNE) will be able to provide valuable and complementary information
about flavor and lepton number when combined with water Cerenkov detectors.
Together, the large number of events and high energy resolution available from
the current suite of neutrino experiments will put much more stringent
constraints on the interior properties of \PNS{s} when the next galactic \CCSN{}
is detected.

\index{diffuse supernova neutrino background}
It is also possible that current neutrino detectors with upgrades or next
generation neutrino detectors will be able to observe the diffuse background of
neutrinos produced by \CCSN{e} over the lifetime of the universe
\citep{Horiuchi:09}.  Predictions for the diffuse MeV scale neutrino background
density depend on the integrated spectrum of neutrinos emitted during \CCSN{e},
especially during the \PNS{} cooling phase \citep{Nakazato:15}.  

\subsection{Impact on \CCSN{} Nucleosynthesis}
\label{sec:nucsyn} 

\index{neutrino driven wind}
Neutrinos can alter the composition of material that is ejected from \CCSN{}{e},
mainly in the innermost ejected regions \citep{Woosley:02}.  
The ejecta most affected by neutrinos is the material that comes from the
surface of the \PNS{}.  Neutrino energy deposition in the atmosphere of the
\PNS{} can provide enough energy to unbind material from the surface and produce
a neutrino driven wind \citep{Duncan:86}.  Once outflowing material reaches large
radii, the temperature drops and heavy nuclei form in the wind.  The heating is
driven by the charged current reactions 
\begin{eqnarray*} 
\nu_e + n &\rightarrow& e^- + p \,\, \textrm{and} \\ 
\bar \nu_e + p &\rightarrow& e^+ + n \nonumber. 
\end{eqnarray*} 
So, in addition to energizing the ejected material,
neutrino interactions change its composition.  The gravitational binding energy
of a baryon at the surface of the \PNS{} is $G M_{\rm pns}m_N/r_{\rm pns}\approx
160 {\rm MeV}$, so a baryon must undergo between ten and fifteen neutrino
captures (given the expected average neutrino energies discussed above) to
escape the potential well of the \PNS{}.  This number of interactions is large
enough to push the material to a composition were electron neutrino capture
balances electron antineutrino capture, which results in an electron fraction
\citep{Qian:96}  
\begin{equation} 
Y_{e,{\rm NDW}} = 
\frac{\lambda_{\nu_e}}{\lambda_{\nu_e} + \lambda_{\bar \nu_e}}.
\end{equation} 
The neutrino capture rates are proportional to $\lambda_{\nu_e} \propto \dot
N_{\nu_e} \langle(\epsilon_{\nu_e}+\Delta_{np})^2\rangle / r^2$ and
$\lambda_{\bar \nu_e} \propto \dot N_{\bar \nu_e}F^N_{\bar \nu_e}
\langle(\epsilon_{\nu_e}-\Delta_{np})^2\rangle / r^2$, where $\Delta_{np} =
1.293 \, {\rm MeV}$ is the neutron proton rest mass difference.    Which heavy
nuclei form depends very strongly on the electron fraction--as well as the
entropy and dynamical timescale--of the outflowing material \citep{Woosley:94,
Sumiyoshi:95, Hudepohl:10, Fischer:10, Roberts:12b, Nakazato:13, Arcones:13}.
Therefore, the final composition depends on the difference between the average
energies of electron neutrinos and antineutrinos.  The magnitude of this
difference in numerical models of \PNS{} cooling is sensitive to the mean field
corrections discussed in section \ref{sec:opacities}, as well as to the method
of neutrino transport \citep{Hudepohl:10, Fischer:10}.  It seems that the wind is
at most marginally neutron rich, but this result depends on the
properties of the assumed nuclear \EOS{} \citep{Roberts:12, Hempel:15,
Mirizzi:15}.   

\index{$\nu$-process}
The $\nu$-process is another mechanism by which \PNS{} neutrinos can alter the
composition of the material ejected from \CCSN{e} \citep{Woosley:90}.  Here,
unlike in the neutrino driven wind, both charged current and neutral current
neutrino interactions are responsible for altering the composition of the
material.  Therefore, neutrinos of all flavors contribute to the process.
Essentially, the $\nu$-process alters the composition of ejected stellar
material by the reactions 
\begin{eqnarray}
(Z,N) + \nu \rightarrow (Z,N)^* + \nu' &\rightarrow& (Z,N-1) + n + \nu'
\nonumber\\
&\rightarrow& (Z-1,N)+p+\nu'  
\nonumber\\
&\rightarrow& (Z-2,N-2)+\alpha+\nu',
\nonumber\\
(Z,N) + \nu_e  &\rightarrow& (Z+1,N-1) + e^{-}, \, \textrm{and}
\nonumber\\
(Z,N) + \bar \nu_e  &\rightarrow& (Z-1,N+1) + e^{+},
\end{eqnarray} 
where $(Z,N)$ denotes a nucleus containing $Z$ protons and $N$ neutrons and
$(Z,N)^*$ denotes an excitepd state of that nucleus.  This is likely responsible
for the production rare isotopes, such as $^{11}$B, $^{19}$F, $^{15}$N
$^{138}$La, and $^{180}$Ta, from much more common nuclei in the envelope of the
\SN{}.  Additionally, it has been suggested that neutrons produced by neutrino
interactions in the helium shell can be rapidly captured on pre-existing heavy
nuclei and make the r-process \citep{Epstein:88}, although later work has shown
it is challenging to achieve the requisite conditions for this process
\citep{Banerjee:11}.  The nuclear yields produced by these process
are sensitive to the time integrated flux and spectrum of neutrinos incident on
the exterior material \citep{Heger:05}.  Therefore, the properties of \PNS{}
neutrino emission, along with neutrino oscillations above the \PNS{}, are very
important to determining the results of $\nu$-process nucleosynthesis
\citep{Heger:05, Banerjee:11}.

For further details on the impact of neutrinos on \CCSN{e} nucleosynthesis, see
the chapters ``Effect of neutrinos on the ejecta composition of core collapse
supernovae'' and ``Production of r-process elements in core collapse
supernovae.'' 

\begin{acknowledgement} 
We acknowledge our collaborators on this subject, including Gang Shen, Vincenzo 
Cirigliano, Jose Pons, Stan Woosley, and Ermal Rrapaj.  LR was supported by NASA 
through an Einstein Postdoctoral Fellowship grant numbered PF3-140114 awarded by 
the Chandra X-ray Center, which is operated by the Smithsonian Astrophysical 
Observatory for NASA under contract NAS8-03060. SR was supported by the US
Dept. of Energy Grant No. DE-FG02-00ER41132. 
\end{acknowledgement}

\end{document}